\documentstyle[aps,preprint,epsf]{revtex}

\begin{document}
\draft

\title{
Synchronous phase clustering in a network of neurons with
spatially decaying excitatory coupling
}
\author{Yuqing Wang$^{1,2}$,  Z. D. Wang$^{1*}$,
Y. -X. Li$^{2}$, and X. Pei$^{3\dagger}, $
}
\address{
$^{1}$ Department of Physics, University of Hong Kong, 
Pokfulam Road, Hong Kong, P.R. China\\ 
$^{2}$ Departments of Mathematics and Zoology, University of British Columbia, 
Vancouver, BC, Canada V6T 1Z2\\
$^{3}$ Center for Neurodynamics, University of Missouri at St. Louis, Missouri 63121, USA
}

\maketitle

\begin{abstract}
Synchronization is studied in a spatially-distributed
network of weekly-coupled, excitatory neurons of Hodgkin-Huxley type. 
All neurons are coupled to each other synaptically 
with a fixed time delay and a coupling strength inversely proportional to the
distance between two neurons. We found that 
a robust, noise-resistant phase clustering state occurred 
regardless of the initial
phase distribution. This has not been shown in previous studies where
similar clustering states were found only when the coupling was inhibitory.
The spatial distribution of neurons in each synchronous cluster is determined
by the spatial distribution of the coupling strength.
Phase-interaction properties of the model neurons in the network are used to
explain why can such a clustering state be robust.
\end{abstract}


\newpage 

Synchronization of coupled neural networks has attracted much interest.
In many parts of the central nervous system (CNS), 
synchronous oscillations have been observed  and are assumed to be correlated to 
specific behaviors, cognitive tasks as well as pathological states\cite{R3a}.
Synchronization in coupled neural networks has been studied theoretically by using various neuronal
models \cite{R4}.
It is generally assumed that neurons in the network are pulse coupled since real neurons
interact with each other only when they generate spikes of action potentials.
The occurrence of synchrony in a network of neurons is 
largely determined
by: (1) the phase-response nature of each neuron, (2) the nature of coupling
(e.g. excitatory or inhibitory) between neurons, (3) the time delay involved in the coupling.
Furthermore, the presence of noise can also influence the 
stability of synchronized states. The fact that neurons distributed across a wide
spatial range can still synchronize \cite{Rn5} poses an important question
as to what makes it possible.
Many studies have indicated that synchronous states can be observed 
when time-delayed synaptic coupling is predominantly 
inhibitory \cite{R5b,R5c,R5d}.
The influence of other important features such heterogeneity and sparseness in the coupling
have also been investigated\cite{R6}.
However, conditions for the occurrence of robust, fully or partially synchronized states
in a network of synaptically coupled excitatory neurons 
have not been fully investigated.

In this Letter, we address this problem with a large network
of excitatory neurons that are organized in a two dimensional square lattice in space.
We found that, in the presence of independent local noise, 
a robust synchronous clustering state can occur regardless of the initial phase distribution.
This state is characterized by the clustering of neurons into
15 synchronized clusters that fire consecutively at a fixed time interval
that is one 15th of their intrinsic period 
\cite{R5b,R5c,Rn9a}.
In previous studies, similar clustering states were found robust only with 
inhibitory mean-field coupling\cite{R5b,R5c}. 
It is known that excitatory coupling usually causes instability to
synchronized state. The occurrence of robust synchronized clusters in the network of
excitatory neurons seems to contradict this well established fact.
Our analysis provides an intuitive explanation of how this is possible.

Figure 1 is a schematic diagram of the network consisting of
an array of 30 by 30 neurons. This network was originally proposed in an attempt to model the network of
an array of electroreceptor cells that are distributed on the rostrum of the paddlefish,
{\em Polydon Daphnia}. The distance between nearest neighbors is taken as the unit length.
The coupling strength between any pair of neurons is inversely proportional to
the distance $R$ between them\cite{R11aa}. The neurons are all identical and are 
coupled to each other through excitatory synapses modeled
by an alpha function with a fixed time delay.
The model that we use to describe each neuron in the network
was previously developed for the electroreceptor cells in the 
paddlefish\cite{R11a}, \cite{R12}, and \cite{R12a}.
It is a modified Hodgkin-Huxley type model involving four ionic currents and a leak current.
\begin{eqnarray}
C_{M}\frac{dV}{dt} &=& -I _{leak}-I_{Na\_ spike}-I_{K\_ spike}-I_{Na\_ slow}
 -I_{K\_ slow}+I_{syn}+ \sqrt{2D}\eta (t)
\end{eqnarray}
where all the ion currents are expressed in terms of ohm's law $I_x=g_x(V-V_x)$ (subscript
``x'' specifies each specific ion current involved). Detailed
expressions for $g_x$, the gating equations, and the parameter values can be found in
\cite{R11a}, \cite{R12}, and \cite{R12a}. 
Independent Gaussian white noise was added to each neurons with uniform
intensity $D$ as described by the last term in Eq.~(1). 
Synaptic coupling was excitatory and
was modeled by an $\alpha $ function 
\cite{R12b} given in the following expression.
\begin{equation}
I_{syn}^{i}=\sum_{j=1}^{N} \frac{W}{R_{ij}} \int\limits_{0}^{\infty }(s/\tau^{2})exp(-s/\tau )S_{j}(t-s)ds
\end{equation}
where  $S_{j}(t)=\sum_{k}\delta (t-t_{k})$ represents the
output spike train from the $\sl j$th neuron firing at times $t_{k}$. 
The delay between the peak time of the presynaptic
pulse and that of the postsynaptic current is fixed at $\tau=7$ ms. 
$R_{ij}$ denotes the distance between the $\sl i$th
and the $\sl j$th neuron (Fig. 1). 
$W (>0)$ is a constant denoting the coupling strength.
For the parameter values used in this model\cite{R12},
each uncoupled neuron exhibits a stable limit cycle oscillation.

In the absence of noise,  
the system can evolve into a state of multiple clusters with carefully
selected initial phase distributions.
For example, if the initial phase distribution 
was uniform, a synchronous phase clustering state of 16 clusters can occur.
The stroboscopic picture of phases for all 900 neurons
in Fig.~2(a) shows how the clusters emerge as time evolves. 
Each neuron fires periodically
while the 16 clusters fire consecutively
at a fixed phase difference. This makes the cycling frequency 
of the averaged electrical activity of the network 16 times
the frequency of each neuron.

In the presence of weak noise ($D=10^{-5}$), 
a robust phase clustering state with 15 clusters was obtained.
Furthermore, the same 15-cluster state 
was always obtained regardless
of the initial phase distribution\cite{R13}.
All other clustering states that occurred in the absence of noise with selected 
initial phase distributions are no longer robust when noise was introduced.
With noise, the same initial 
phase distribution as in Fig.~2(a) evolved into a clustering state of 15 clusters
(Fig.~2(b)).
Simulations further indicated that this 15-cluster state 
was robust in the presence of weak noise.
Starting with a fully in-phase-synchronized initial state where each neuron
fires simultaneously (Fig.~2(c), the
synchronization was rapidly destroyed and the 15-cluster 
state eventually emerged. Had there been no noise, 
this fully in-phase-synchronized state would have remained unchanged 
for the same time duration.
Similarly, when the system was initiated at a 2-cluster
state (Fig.~2(d)) which would persisted in the absence
of noise was destroyed by the noise and eventually evolved into the 15-cluster
state, although a longer evolution time
was required.  We tried many different initial phase distributions and found that
the same 15-cluster state emerged all the time,
although the transients and the detailed distribution 
of neurons in each cluster were different.

A typical pattern of spatial distribution of neurons in each cluster is
shown in Fig.~3 that corresponds to the 15-cluster state reached in Fig.~2(b).
Three important features of such distributions were found:
(1) the number of neurons in each cluster was approximately the same, i.e. about 60 (=900/15);
(2) neurons in each synchronized cluster were almost evenly distributed in space;
(3) the even distribution of neurons in each cluster also indicated that
the average inter-cluster distance between neurons is similar between
different clusters.

The results presented here
are different from previous studies in networks with mean-field
type excitatory coupling where phase clustering states cannot persist in the
presence of week noise (see Fig. 7 in Ref. \cite{R5c} and \cite{R14}).
Our network differs from the mean-field type of coupling because the coupling
strength is inversely proportional to the distance between two neurons.
The question is how this difference makes the robust, noise-resistant
clustering state possible. Furthermore, why is the 15-cluster state the 
asymptotically stable one? What is special about the fixed phase difference
$1/15\approx 0.067$? Here are the conclusions reached by the following
analysis:  the fixed phase difference and
the number of clusters are determined by the phase-interaction properties between
two neurons while the robustness of the clustering state
is achieved because the neurons in the network self-organizes into
clusters in such way that the stabilizing inter-cluster 
interactions are stronger than the destabilizing
intra-cluster interactions.

Several important features of the 15-cluster state 
can be explained by the phase-interaction properties between two neurons.
Phase-interaction between two neurons depends on how the neurons are coupled to
each other and how each neuron responds to interactions between them.
These properties are characterized mathematically 
by the so called interaction function that is defined
as the convolution between the synaptic current $i_{syn}(t)=\frac{R}{W}I_{syn}(t)$
and the adjoint or the phase-response function
$Z(t)$ \cite{R12e} over one oscillation period.
\begin{equation}
H(\phi)=\frac{1}{T}\int\limits_{0}^{T} dt Z(t)
 \int\limits_{0}^{\infty }(s/\tau^{2})exp(-s/\tau )\sum_{k=-\infty}^{0}\delta (t-s+kT-\phi)ds.
\end{equation}
where $\phi$ is the phase difference between the two oscillators that satisfies
\begin{equation}
\frac{d \phi }{dt}=-wg(\phi )
\end{equation}
where $g(\phi)=H(\phi)-H(-\phi)$ is the odd part of $H(\phi)$ 
and $w=W/R$ is the coupling strength between the two neurons.
Thus, the phase-lock solutions are the zeros of $g(\phi)$ and the stability
is determined by the sign of $g'(\phi)$ (stable if $g'>0$, unstable if 
$g'<0$)\cite{R12c,R12d}. The $g(\phi)$ and $g'(\phi)$ calculated with the model described by
Eqs.~(1-2) and the synaptic current given by Eq.~(2) are shown
in Fig.~4(a) for the $\tau$ value used in Figs.~2-3. There are 7 zeros in one complete
period of which 3 pairs are symmetric. Thus only 4 possible phase-locking states are possible:
in-phase ($\phi=0$) that is unstable, anti-phase ($\phi=0.5$) that is stable,
a fixed phase difference ($\phi\approx 0.062$) that is stable, and an unstable
phase-locking state. The $g'(\phi)$ curve shows that the anti-phase solution is 
weekly stable but the phase-lock solution at $\phi=0.062$ is strongly stable. 
Now we see that the number of clusters and this strongly
stable phase-locking state is closely related 
since $0.062$ is close to the actual phase difference
(0.067) in the 15-cluster state. 
Fig.~4(b) shows that the locked phase difference
of this phase-locking
state increases as the synaptic time delay $\tau$ increases. This implies that
at larger values of $\tau$, the number of clusters in the robust clustering state
should decrease. 
However, this simple explanation does not answer how could the neurons
in each synchronized cluster maintain their stability and why the 16-cluster
state was not the asymptotic state 
although $1/0.062\approx 16.13$ is closer 16 or even 17 but not 15. 

To further demonstrate the robustness of the 15-cluster state we need to study the
stability of the clustering state based on the phase description of coupled neural
network. With week coupling, each neuron can be approximately described
by their respective phase variation $\phi_i$
resulting from synaptic interactions with other neurons.
Such interactions are characterized by the same interaction function $H$ calculated
in Eq.~(3). Thus, the network can be described by the following 900 phase coupled
equations.

\begin{equation}
\frac{d \phi _{i} }{dt}=\frac{1}{N}\sum\limits_{j=1 (j\neq i)}^{N} w_{i,j}H(\phi _{j}-\phi _{i}).
\end{equation}
where $w_{i,j}=w_{j,i}=\frac{W}{R_{i,j}}$ is the coupling strength between the
$i$th and the $j$th neurons.
It is hard to use these equations to determine all the possible solutions.
However, they can be useful in determining the stability of a known solution.
Instead of solving the eigenvalue problem of the huge Jacobian matrix,
we here focus on the stability of the clustering state
when the phase of a single neuron is perturbed. For simplicity, we
assume that the average distances between two neurons within and between all 
clusters are identical (confirmed by our simulations), 
we can derive the equation
that approximately describes the time evolution of the perturbation of any neuron
$\delta \phi=\phi_i-\Phi_i$ (where $\Phi_i$ is the phase of all neurons
in the cluster where neuron $i$ belongs).

\begin{equation}
\frac{d (\delta \phi) }{dt}
\approx -[\bar{w_0} g'(0)+ \bar{w_1}\sum\limits_{j=1}^{n-1} g'(j/n)] \delta \phi,
\end{equation}
where the intra-cluster coupling strength,
$\bar{w_0}=\frac{1}{n} \sum_{i=0}^{n-1} \frac{2}{n_i(n_{i}-1)}\sum_{i',j'=1,i'\neq
j'}^{n_i}
\frac{W}{R_{i',j'}}$,
is the average coupling strength between two neurons within one cluster.
The inter-cluster coupling strength, 
$\bar{w_1}=\frac{2}{n(n-1)}\sum_{i,j=1 (i\neq j)}^{n-1}\frac{1}{n_i n_j} 
\sum_{i'=1}^{n_i} \sum_{j'=1}^{n_{j}} \frac{W}{R_{i',j'}}$ is the average coupling strength
between two neurons in different clusters. 
In these expressions, $n$ is the number of clusters and $n_i$ 
is the number of neurons in the $i$th cluster.

Equation (6) is valid for all neurons in the network.
Therefore, the stability of a particular clustering state is achieved when
$\lambda=\bar{w_0} g'(0)+ \bar{w_1}\sum_{j=1}^{n-1} g'(j/n)>0$ and the most robust
clustering state is the one that maximizes $\lambda$.
$g'(0)<0$ is generally true for excitatory coupling (see Fig.~4(a)). 
Thus intra-cluster interactions destabilize synchrony in each cluster,
consistent with the known effect of excitatory coupling.
However, the cluster can still be stable since
$g'(j/n)$ ($j=1,\cdots,n-1$) can be positive (see Fig.~4(a)). For a given number
of clusters $n$, if $g_n=\sum_{j=1}^{n-1} g'(j/n)>0$, the n-cluster state is stable if
$\bar{w_1}/\bar{w_0} > |g'(0)|/g_n$ is satisfied. The most robust clustering state
is the one with $n^*$ clusters where $n^*$ maximizes $\lambda$.
In the present network $n^*=15$.

The analysis indicates that the stability of multi-cluster states
with excitatory coupling is favored by two conditions:
(1) $g'(\phi)$ is strongly positive and/or positive for a larger part of the cycle (see Fig.~4(a));
(2) $\bar{w_1} > \bar{w_0}$ (e.g. $\bar{w_1},\bar{w_0}\approx
9.5e^{-4}, 9.1e^{-4}$ in Figs.~2-3), 
i.e. a larger average inter-cluster coupling than
average intra-cluster coupling.
Both conditions are met by our network.
Fig.~4(a) shows that
our $g'(\phi)$ is indeed positive over a large part of the cycle and has two large
positive peaks located at $\phi=0.062$ and $1-0.062$. The fact
that $|g'(0.062)|>>|g'(0)|$ not only explains why
the asymptotic state has as many as 15 clusters but also indicates that this
clustering state could still be robust even if condition (2) were not met.
Spatial decay in coupling strength is crucial for the network to satisfy
condition (2). It allows
the network to minimize $\bar{w_0}$ by scattering the
neurons in each cluster far apart from each other. This explains why
neurons are evenly distributed in each cluster (Fig.~3).

This analysis shows that when the coupling strength is allowed to differ
between different neurons, robust clustering states can occur in networks
of excitatory neurons. This result is not model specific, although the number
of clusters depends on specific features of the model and some key parameters 
such as the delay time $\tau$. This results is further supported by numerical
studies of other neuronal models including the integrate-and-fire model 
(work in progress), provided the conditions outlined above are satisfied.

The fact that distributed coupling strength
can generate robust synchronous clusters 
may have some relevance in understanding the 
dynamical variations of brain functions.
First, neurons can join different synchronized
clusters depending on how the pattern of coupling strength is modified.
It is possible that these patterns can be mapped
to the functional states of the brain. The patterns of synchronized clusters may be 
used as codes in some cognitive tasks.
Second,  associative memory
patterns could be stored in the patterns of coupling strength and the clustering patterns
could be `selected' by the coupling pattern.
Furthermore, these patterns could be modified during learning processes.
Such memory patterns can be retrieved once the 
clustering patterns are retrieved through dynamical evolution. 
Finally,  the large number of clusters ($n$) in the clustering states makes the average
electrical activity of the network oscillate with a frequency $n$ times
the intrinsic frequency of each individual neuron.
It is worth noting that a high-frequency
network oscillation (200Hz,  several times
the spontaneous firing frequency of a single neuron) was reported
in experiments on pyramidal cells in the CA1 hippocampal region of rats \cite{R11}. 

The work was supported by an HKU{ \& }IBM supercomputing program. 
We thank Prof. Robert M. Miura for his critical reading of the manuscript.
Y. Wang acknowledges support from The Pacific Institute for Mathematical Sciences.

* To whom correspondence should be addressed. Email address: zwang@hkucc.hku.hk \\ 
$\dagger$ Present Address: St. Jude Medical, Sylmar, CA 91342, USA.

\begin{figure}[tbp]
\epsfbox{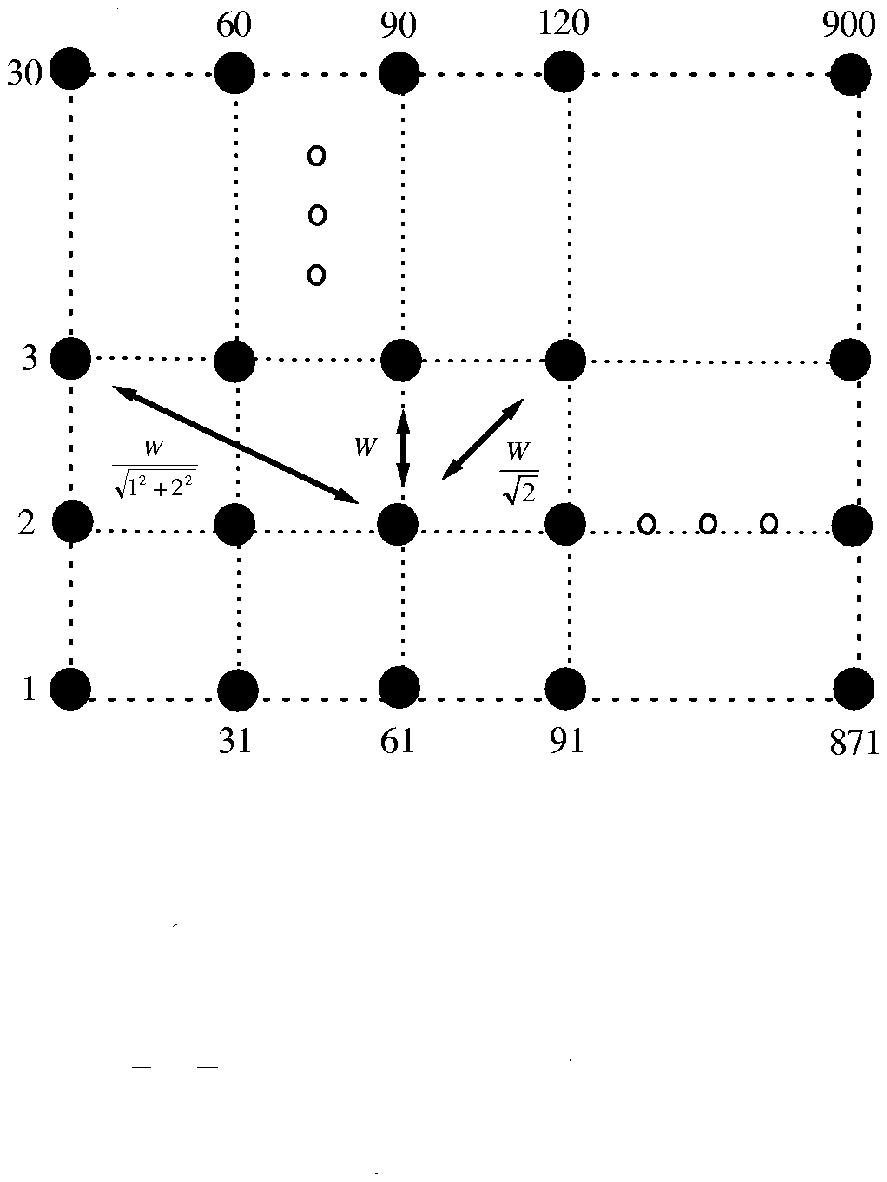}
\caption{ Schematic diagram of  a  two-dimensional lattice 
of $30\times 30$ modified HH neurons, with the black dots
representing the neurons (1 to 900), where the coupling strength
of the $i$th and $j$th neurons is $W/R_{ij}$ with $R_{ij}$ as the distance
between neurons $i$ and $j$.
}
\end{figure}


\begin{figure}[tbp]

\epsfbox{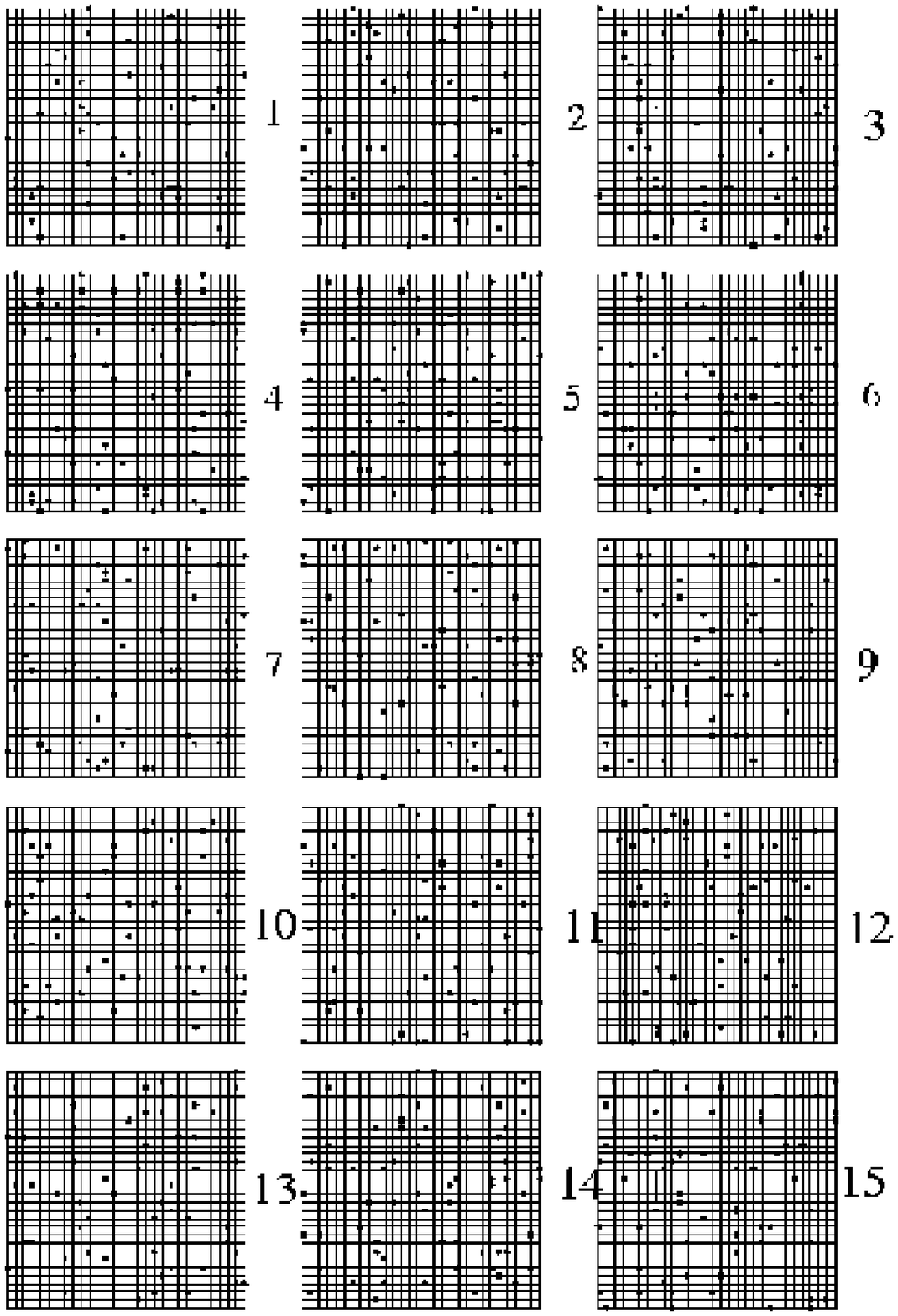}



\caption{Typical spatial distribution of neurons in each cluster for the 15-cluster state in Fig. 2(b).
The synchronized pattern labeled $1$ to
$15$ appear consecutively as time evolves.
}
\end{figure}

\begin{figure}[tbp]

\epsfbox{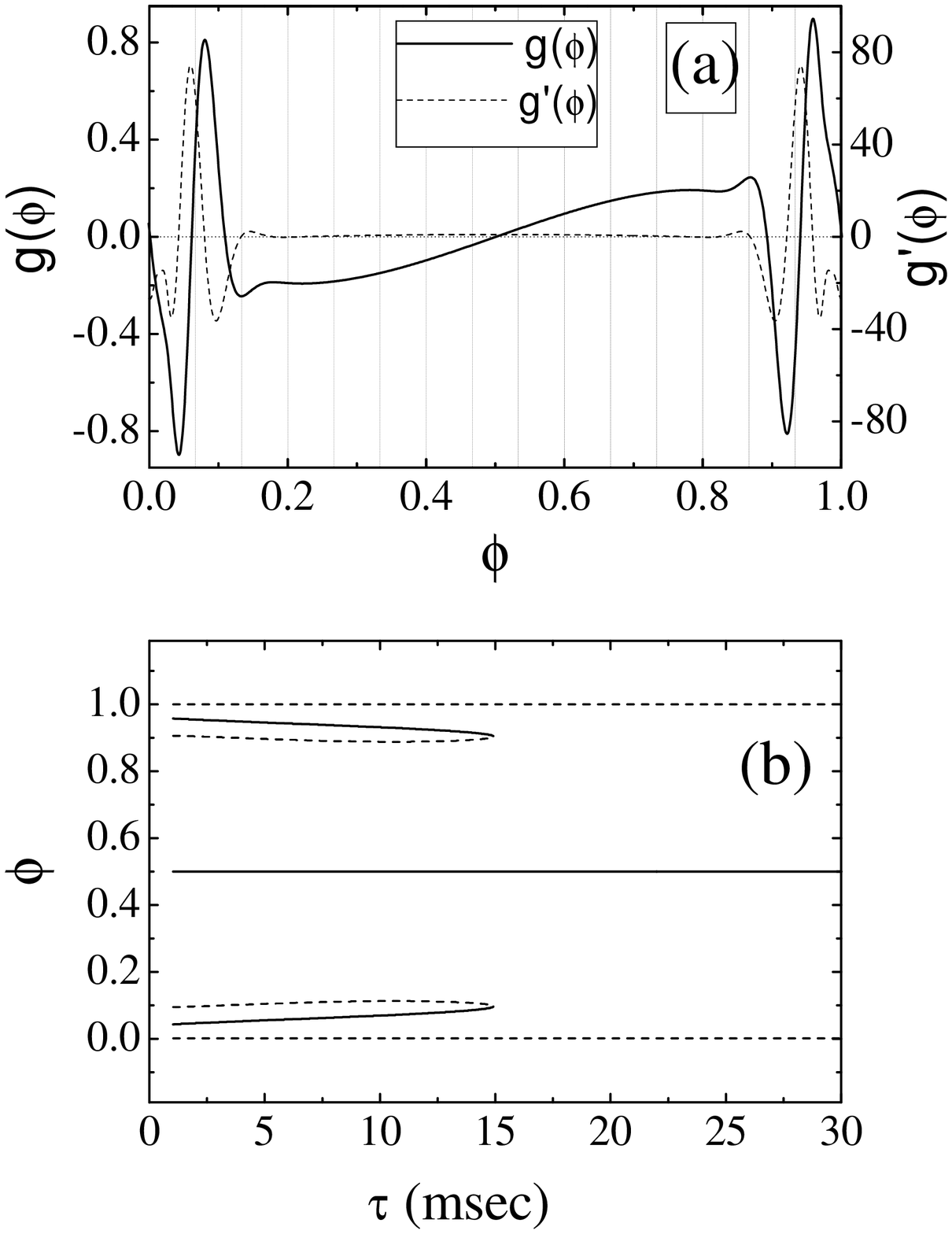}



\caption{
(a) $g(\phi )$ (solid line) and $g'(\phi )$ (dashed line)
plotted as a function of the phase $\phi $ when $\tau =\ 7ms.$ The scale of
$g(\phi )$ is shown on the left side of the figure and that for $g'(\phi )$ on the right side. 
(b) Phase difference of stable (solid) and unstable (dashed) states plotted as a function
of $\tau$. 
}
\end{figure}


\begin{references}
\bibitem{R11aa} A coupling strength $\propto 1/R^{2}$ was also tried. The conclusion regarding synchronous
phase clustering is essentially the same as that for $1/R$.
\bibitem{R12a} In this paper, the parameters in the equation are the same
as those in \cite{R11a} and \cite{R12}, which were determined from experimental 
data on peripheral cold receptors.
\bibitem{R13} The intensity of noise chosen here ($D=1\times 10^{-5}$) is weak but not
                  negligible. 
              The phenomenon discussed here is preserved even when the noise 
               intensity is as high as $D=1\times 10^{-4}$( about $10\%$ of  
                $<I_{syn}>$, the average value of the coupling term).
\bibitem{R14} If we employ uniform inhibitory or excitatory coupling, 
the main results are the same as those in Ref. \cite{R5b,R5c}, where 
an integrate-and fire  model and an HH model are employed.
\end{references}
\end{document}